\documentclass[preprint,amsfonts,amsmath,amssymb,eqsecnum,aps,showpacs,showkeys,floatfix]{revtex4}
\newcommand{\N}{\mathbb{N}}

\newcommand{\D}{\not\!\!{D}}

\newcommand{\bs}[1]{\mathbf{#1}}
\newcommand{\Bold}[1]{\mbox{\boldmath$\mathit{#1}$}}
\newcommand{\sBold}[1]{\mbox{\boldmath$\mathit{\scriptstyle{#1}}$}}
\newcommand{\ssBold}[1]{\mbox{\boldmath$\mathit{\scriptscriptstyle{#1}}$}}
\newcommand{\C}{\mathbb{C}}

\begin{document}

\title{\textbf{Charges and Coupling Strengths
 in Gauge Theories\\ with Direct Product Symmetry Groups}}
\author{J. LaChapelle}
%\affiliation{1387 NW Ashley Dr, Albany, OR 97321\\jlachapelle@attbi.com\\541-791-2348}

%%% ----------------------------------------------------------------

%%% ----------------------------------------------------------------

\begin{abstract}
For gauge theories with direct product internal symmetry groups,
the relationship between internal quantum numbers (charges) and interaction
coupling strengths is examined. In these types of theories, the
Lagrangian density may contain non-trivial factors multiplying the
matter field terms, and these factors can modify the interaction
coupling strengths i.e., the gauge/matter field vertex factors.
Consequently, a matter field can carry a given internal charge yet
couple to the associated gauge field with an apparent fractional
charge. An example with $SU(3)\otimes U(2)$ symmetry is presented
in which the matter fields can have integer $U(2)$ charges but
fractional $U(2)$ coupling strengths.

\end{abstract}



\maketitle

\section{Introduction}

This paper studies the relationship between internal quantum
numbers---commonly called charges---and coupling strengths
in gauge theories with direct product Lie groups. For the most
part, the analysis reproduces what is already known since gauge
field theory is, by now, well studied and understood (see e.g.
\cite{HEN}-\cite{ORA}). However, the analysis uncovers two
important points that may be of relevance in physical
applications.

The first point is that the group theory mathematics (see e.g.
\cite{COR}, \cite{GIL}) suggests the matter field irreducible
representations (irreps) for the direct product group include
\emph{all} combinations of irreps of the component subgroups. The
second point is that matter field terms in the Lagrangian density
can contain non-trivial multiplicative factors that do not destroy
the salient elements of the gauge theory. Taken together, these
two results imply the possibility of scaled gauge/matter field
vertex couplings. In particular, it
implies the possibility of matter fields with integer charges but
fractional coupling strengths.

Our analysis starts with the Lie algebra of a direct product
group. There exists a special Lie algebra basis that implies
neutral and paired, oppositely-charged gauge bosons. This basis
corresponds to `physical'\footnote{We use the qualifiers ` '
around physical because, although for a broken symmetry the
physical gauge bosons correspond to a definite basis in the Lie
algebra, for an unbroken symmetry all bases are equivalent by
virtue of the local symmetry. On the other hand, once one chooses
a matter field representation, a natural choice of Lie algebra
basis for unbroken symmetry corresponds to gauge bosons that
exchange a quantum of charge with the matter fields. This special
basis can be regarded as `physical'.} gauge bosons and
characterizes the gauge bosons' intrinsic quantum numbers.
Likewise, one can choose an associated basis in the vector space
furnishing a faithful irrep of the direct product group. Relative
to this associated basis, the eigenvalues of the diagonal Lie
algebra elements characterize the intrinsic quantum numbers of the
elementary matter fields. Thus \emph{the kinematical structure
encodes the intrinsic quantum numbers of the elementary gauge and
matter fields}.

On the other hand, the gauge/gauge and gauge/matter field
couplings are characterized by their respective covariant
derivatives. There is enough scale ambiguity in the Lagrangian
density to allow the identification of the eigenvalues of the
neutral conserved charge operators with the intrinsic quantum
numbers---thereby ensuring the fields appearing in the Lagrangian
density are elementary fields \cite{WEI}. The two sets of
parameters---so identified---will be referred to as
\emph{intrinsic charges}. The (properly scaled) intrinsic charges
multiply vertex factors in Feynman diagrams and contribute to the
gauge/matter field coupling strengths. However, this is not the
sole contribution to the coupling strengths.

The Lagrangian density admits a multiplicative factor for each
matter field representation. For direct product groups, the
representations may be related in such a way that the factors are
not trivial in the sense that the factors cannot all be absorbed into the matter field definition.
Consequently, non-trivial factors may multiply the intrinsic charges in the
conserved currents, and, therefore, affect the gauge/matter field
vertex factors (or coupling strengths). These renormalized
coupling strengths will be referred to as \emph{extrinsic charges}
since they are actual measured quantities; provided the associated
particle is observable.

\emph{The presence of non-trivial multiplicative factors implies
an inequality between the associated intrinsic and extrinsic
charges of elementary matter fields}. Although it is the extrinsic
charge that is observed, the notion of intrinsic charge is
theoretically useful. An example is presented in section \ref{sec. example} that exhibits matter
fields with integer $SU(3)\otimes U(2)$ intrinsic charges---yet
they couple to $U(2)$ gauge bosons with fractional extrinsic
charges.

\section{Intrinsic vs. Extrinsic Charges}
There are at least two starting points for characterizing quantum
numbers of elementary particles and fields. One is kinematical and
the other dynamical in nature.

The kinematical starting point stems from viewing an elementary
particle state as furnishing a faithful irrep\footnote{The
assumption that the state is elementary dictates an irreducible
representation since otherwise the state would have subcomponents
that transformed irreducibly. These subcomponents would then be
considered ``elementary''. Faithful representations are stipulated
so that all aspects of the abstract symmetry group are carried by
the representation.} of some assumed internal symmetry group that
commutes with the Poincare group. The representations are labelled
by certain parameters (that must be scalars by Poincare
invariance) which then serve to characterize physical properties
(apart from momentum and spin/helicity) of the elementary particle
state. We will refer to these parameters as kinematical (internal)
quantum numbers.

The other starting point is the Lagrangian---hence dynamical. By
way of Noether's theorem, symmetries of the Lagrangian lead to
conserved currents that in turn lead to time independent quantum
charge operators. Their equal-time commutators with the field
variables of the Lagrangian yield what we will refer to as
dynamical (internal) quantum numbers. Importantly, scale ambiguity in the
Lagrangian can be exploited to guarantee equality between the
kinematical and dynamical quantum numbers.

\subsection{Kinematical quantum numbers}\label{sec. kinematics}

Given that a physical system is invariant under some internal
symmetry group, it is possible to deduce some general properties
or attributes of the associated gauge and matter fields based
solely on the mathematics of the symmetry group and its
representations \cite{COR}, \cite{GIL}. In particular, the mathematics identifies special
bases and associated eigenvalues (quantum numbers) in the vector
spaces furnishing the representations. For local symmetries, these
special bases can be chosen at each spacetime point, essentially
creating an unchanging structure by which to associate the
unchanging internal quantum numbers of elementary particles.

We begin with a gauge field theory with an internal symmetry group
that is a direct product group $G=G_1\otimes\cdots\otimes
G_n=:\otimes\,G_n$ where $n\in\N$ and the $G_i$ with
$i\in\{1,\ldots,n\}$ are Lie groups that mutually commute.
Associated with each subgroup $G_i$ is a Lie algebra
$\mathcal{G}_i$ with basis
$\{\bs{g}_{a_{i}}\}_{a_i=1}^{\mathrm{dim}G_i}$. The full Lie
algebra is $\mathcal{G}:=\oplus \mathcal{G}_n$ (in obvious
notation). Recall that the Lie algebra does not uniquely determine
the Lie group.

Consider the adjoint representation $ad:\mathcal{G}_i\rightarrow
GL(\mathcal{G}_i)$ of the complex extension of $\mathcal{G}_i$ on
$\mathcal{G}_i$. For a given element $c^{a_i}\bs{g}_{a_i}$ (with
$c^{a_i}\in \C$) in the Lie algebra, the adjoint representation
yields a secular equation
\begin{equation}\label{secular}
  \prod_{k_i=0}^{r_i}(\lambda-\alpha_{k_i})^{d_{k_i}}=0
\end{equation}
where the $\alpha_{k_i}$ are the (complex) roots of the secular
equation with multiplicity $d_{k_i}$. Since $\lambda=0$ is always
a solution, we put $\alpha_0=0$. Note that
$\sum_{k_i=0}^{r_i}d_{k_i}=\mathrm{dim}G_i$. Associated with the
roots $\alpha_{k_i}$ (which may not all be distinct in general)
are $r_i$ independent eigenvectors.

The roots and their associated eigenvectors determine the
well-known Jordan block form of the element
$ad(c^{a_i}\bs{g}_{a_i})$. That is, there exists a non-singular
transformation of $ad(c^{a_i}\bs{g}_{a_i})$ into Jordan canonical
form. With respect to the Jordan canonical form, the vector space
that carries the representation $ad(\mathcal{G}_i)$ (and hence the
Lie algebra) decomposes into a direct sum of subspaces:
\begin{equation}\label{decomposition}
\mathcal{G}_i=\sum_{{\alpha}_{k_i}}\oplus V_{\alpha_{k_i}}
\end{equation}
with each $V_{\alpha_{k_i}}$ containing one eigenvector and
$\mathrm{dim}V_{\alpha_{k_i}}=d_{k_i}$.

The above decomposition is with respect to any given element in
the Lie algebra. Regular elements are defined by the conditions:
(i) that they lead to a decomposition that maximizes the distinct
roots $\alpha_{k_i}$ (equivalently, minimize the dimension of
$V_{\alpha_{k_i}}$), and (ii) they all determine the same
$V_{0_i}$. For decomposition associated with regular elements, the
subspaces $V_{\alpha_{k_i}}$ have potentially useful properties
for describing physical gauge bosons:
\begin{itemize}
\item{$[V_{0_i},V_{0_i}]\subseteq V_{0_i}$ and hence $V_{0_i}$ is a subalgebra.
It is known as a Cartan subalgebra.}
\item{The subspace $V_{0_i}$ carries a representation of the Cartan subalgebra.
Since its rank is $0$, the Cartan subalgebra is solvable; in fact
nilpotent.}
\item{$[V_{0_i},V_{\alpha_{k_i}}]\subseteq V_{\alpha_{k_i}}$. Hence, each
$V_{\alpha_{k_i}}$ is invariant with respect to the action of
$V_{0_i}$ and so carries a representation for $V_{0_i}$. Moreover,
since $V_{0_i}$ is solvable, it has a simultaneous eigenvector
contained in $V_{\alpha_{k_i}}$. More specifically, associated
with the secular equation for an element of the subalgebra
$V_{0_i}$ with basis $\{\bs{h}_{s_i}\}_{s_i=1}^{d_{0_i}}$ is a set
of $d_{0_i}=\mathrm{dim}V_{0_i}$ roots, collectively denoted by
$\Bold{q}_i:=({q_1}_i,\ldots,{q_{d_0}}_i)$, and a corresponding
eigenvector $\bs{e}_{{\alpha}_{k_i}}\in V_{\alpha_{k_i}}$ such
that
\begin{equation}\label{eigenvalue}
  [\bs{h}_{s_i},\bs{e}_{{\alpha}_{k_i}}]=q_{s_i}\bs{e}_{{\alpha}_{k_i}}\;,
\end{equation}
or more succinctly,
\begin{equation}\label{eigenvalues}
 [\bs{h}_i,\bs{e}_{{\alpha}_{k_i}}]=\Bold{q}_{i}\bs{e}_{{\alpha}_{k_i}}\;,
\end{equation}
 In particular, this holds for $\alpha_0=0$. That is, there exists
 an $\bs{e}_{{0}_i}\in V_{0_i}$ such that
\begin{equation}\label{zero}
[\bs{h}_i,\bs{e}_{{0}_i}]=\bs{0}\;.
\end{equation}
 }
\item{If $V_{0_i}$ is contained in the derived algebra of
$\mathcal{G}_i$, then for $V_{\alpha_{k_i}}$, there is at least
one $V_{\beta_{k_i}}$ such that
$[V_{\alpha_{k_i}},V_{\beta_{k_i}}]\subseteq V_{0_{k_i}}$. This
implies that, for  $\Bold{q}_i$ associated with each
$\bs{e}_{\alpha_{k_i}}$, there is at least one
$\bs{e}_{\beta_{k_i}}$ with roots $-\Bold{q}_i$. Additionally, any
$\Bold{q}'_i\neq -\Bold{q}_i$ must be a rational multiple of
$\Bold{q}_i\neq 0$. }
\end{itemize}

These properties can be used to characterize the `physical' gauge
bosons if we make one restriction: for $\alpha_{k_i}\neq 0$,
$\mathrm{dim}\oplus V_{\alpha_{k_i}}=\mathrm{dim}G_i-d_{0_i}=r_i$.
That is $\mathrm{dim}V_{\alpha_{k_i}}=1$ for all $\alpha_{k_i}\neq
0$. Without this restriction, there would be no means
(mathematically) to distinguish between basis elements, and hence
gauge bosons, in a given $V_{\alpha_{k_i}}$. As a consequence of
this restriction, we must have $[V_{0_i},V_{0_i}]=0$ since
otherwise $[[V_{0_i},V_{0_i}],V_{\alpha_{k_i}}]$ in the Jacobi
identity leads to a contradiction.

The commutativity of $V_{0_i}$ is a necessary condition for
$\mathcal{G}_i$, and hence $\mathcal{G}$, to be the direct sum of
one-dimensional abelian and/or simple algebras. Moreover,
eventually the Lie algebra elements will be promoted to quantum
fields so the adjoint carrier space is required to be Hilbert.
Therefore, the inner product on the Lie algebra is required to be
positive definite. This implies the Lie algebra is the direct sum
of $u(1)$ and/or \emph{compact} simple complex algebras. With
this identification, all the results of root space analysis of
compact, semisimple\footnote{Semisimple Lie algebras uniquely
decompose into a direct sum of simple Lie algebras.} Lie algebras
become applicable.

If the symmetry is not broken under quantization, then we can
conclude that the quantized gauge fields associated with the Lie
algebra $\mathcal{G}_i$ describe gauge bosons characterized by the
set of roots $\Bold{q}_i$. We refer to these as \emph{kinematical
(internal) quantum numbers} for the gauge bosons. They correspond
to physical, i.e. measured, properties of the gauge bosons for
broken symmetries, and for unbroken symmetries they are physically
relevant once a choice of matter field representations has been
made.

Evidently, gauge bosons associated with the $\bs{h}_{s_{i}}$ carry
no kinematical `charge' and those associated with the
$\bs{e}_{\alpha_{k_i}}$ carry the $\Bold{q}_{i}$ kinematical
`charges'. Note that $\bs{e}_{-\alpha_{k_i}}$ carries
$-\Bold{q}_{i}$ charges. It is in this sense that the Lie algebra
basis, defined by (properly restricted) decomposition
(\ref{decomposition}), characterizes the physical gauge bosons. It
should be kept in mind that the class of Lie algebras under
consideration are \emph{complex}. However, for the groups of
interest in gauge theory applications, the basis elements are
linearly independent over the complex numbers. Hence, the Lie
algebra can be taken to be \emph{real} (having the same basis as
its complex counterpart) provided the gauge fields are allowed to
be complex.

Turn now to the matter particle states. We will confine our
attention to Dirac spinors.\footnote{We will not display the
spinor components of the matter fields since we work in Minkowski
space-time and they play no (apparent) role in internal
symmetries.} Let $V_{\sBold{R}_i}$ be a vector space that
furnishes a representation of $G_i$ having basis
$\{\Bold{e}_{l_i}^{(\sBold{R}_i)}\}_{l_i=1}^{d_{R_i}
:=\mathrm{dim}V_{\sBold{R}_i}}$. And let
$\rho^{(\sBold{R}_i)}:G_i\rightarrow GL(V_{\sBold{R}_i})$ denote a
faithful irrep. The $\Bold{R}_i$ is a collection
$(R_i^1,\ldots,R_i^{d_{0_i}})$ of $d_{0_i}$ numbers and serves to
label the representation.

Given some set $\{\Bold{R}_i\}$, suppose the corresponding set of
fields $\{\Bold{\Psi}^{(\sBold{R}_i)}\}$ furnish inequivalent
irreps $\rho^{(\sBold{R}_i)}(G_i)$ of the $G_i$. The associated
tensor product representation
\begin{equation*}
\rho^{(\times \sBold{R}_n)}(\otimes G_n
):=\rho^{(\sBold{R}_1)}(G_1)\otimes,\ldots,\otimes\rho^{(\sBold{R}_n)}(G_n)
\end{equation*}
of the direct product group is also irreducible (where $\times
\Bold{R}_n:=(\Bold{R}_1,\ldots,\Bold{R}_n)$ denotes an element in
the cartesian product
$\{\Bold{R}_1\}\times,\ldots,\times\{\Bold{R}_n\}$). In fact for
the class of groups under consideration here, all irreps of $G$
are comprised of all possible combinations of relevant
$\{\Bold{R}_i\}$ \cite{COR}. \emph{That the irreps of
$\rho^{(\times \sBold{R}_n)}(\otimes G_n )$ include all relevant
combinations of component irreps suggests including these
representations in physical models}. The idea is these relevant combinations of irreps can be identified with elementary fields.

The corresponding Lie algebra representation
\begin{equation*}
\rho_e{'{^{(\times \sBold{R}_n)}}}(\oplus
\mathcal{G}_n):={\rho_{e}{'}{^{(\sBold{R}_1)}}}(\mathcal{G}_1)
\oplus,\ldots,\oplus{\rho_{e}{'}{^{(\sBold{R}_n)}}}(\mathcal{G}_n)
\end{equation*}
(where $\rho_e'$ is the derivative map of the representation
evaluated at the identity element) is likewise irreducible for all
combinations of $\{\Bold{R}_i\}$ that are associated with irreps
of the $\mathcal{G}_i$.

The representations $\rho^{(\sBold{R}_i)}(G_i)$ are largely a
matter of choice depending on physical input. By assumption, the
internal degrees of freedom associated with $G_i$ of
\emph{elementary} particles correspond to the basis elements
$\{\Bold{e}_{l_i=1}^{(\sBold{R}_i)}\}_{l_i}^{d_{R_i}}$ spanning
$V_{\sBold{R}_i}$. Hence, a given label $\Bold{R}_i$ (partially)
characterizes the elementary particles (along with Lorentz
labels). In particular, a basis is chosen such that the
representation of the diagonal Lie algebra elements is (no
summation implied)
\begin{equation}\label{full representation}
\rho_{e}{'}{^{(\sBold{R}_i)}}(\bs{h}_{s_i})\Bold{e}_{l_i}^{(\sBold{R}_i)}
=q^{(m)}_{s_i,l_i}\Bold{e}_{l_i}^{(\sBold{R}_i)}\,.
\end{equation}
where $q^{(m)}_{s_i,l_i}$\footnote{The $(m)$ superscript indicates
``matter''.} are $(d_{0_i}\times d_{R_i})$ imaginary numbers since
the Lie algebra generators must be anti-hermitian (which implies
the $\rho_{e}{'}{^{(\sBold{R}_i)}}(\bs{h}_{s_i})$ must also be
anti-hermitian). In an obvious short-hand notation,
\begin{equation}\label{full representation 2}
  \rho_{e}{'}{^{(\sBold{R}_i)}}(\bs{h}_{s_i})\Bold{e}^{(\sBold{R}_i)}
=\Bold{q}^{(m)}_{s_i}\Bold{e}^{(\sBold{R}_i)}\,.
\end{equation}
where
$\Bold{q}^{(m)}_{s_i}:=(q^{(m)}_{s_i,1},\ldots,q^{(m)}_{s_i,d_{R_i}})$
and ${(\Bold{e}^{(\sBold{R}_i)})}^{\mathrm{T}}
=(\Bold{e}_1^{(\sBold{R}_i)},\ldots,\Bold{e}_{d_{R_i}}^{(\sBold{R}_i)})$.
Hence, $\Bold{q}^{(m)}_{s_i}$ can serve to label the basis
elements corresponding to elementary matter particle states for a
given representation labelled by $\Bold{R}_i$. In this sense, the
elementary matter particles carry the \emph{kinematical (internal)
quantum numbers} $\Bold{q}^{(m)}_{s_i}$.

Taking the complex conjugate of (\ref{full representation}), gives
\begin{equation}\label{anti-representation 1}
[\rho_{e}{'}{^{(\sBold{R}_i)}}(\bs{h}_{s_i})]^{*}
{\Bold{e}_{l_i}^{(\sBold{R}_i)}}^{*}
=-q^{(m)}_{s_i,l_i}{\Bold{e}_{l_i}^{(\sBold{R}_i)}}^{*}\;.
\end{equation}
Hence,
\begin{equation}\label{anti-representation 2}
{\Bold{e}^{(\sBold{R}_i)}}^{\dag}
[\rho_{e}{'}{^{(\sBold{R}_i)}}(\bs{h}_{s_i})]^{\dag}
=-\Bold{q}^{(m)}_{s_i}{\Bold{e}^{(\sBold{R}_i)}}^{\dag}\;.
\end{equation}
So $\{{\Bold{e}_{l_i}^{(\sBold{R}_i)}}^{\dag}\}$ furnishes a
conjugate representation of $G_i$ and is obverse to
$\{{\Bold{e}_{l_i}^{(\sBold{R}_i)}}\}$. That is,
$\{{\Bold{e}_{l_i}^{(\sBold{R}_i)}}^{\dag}\}$ represents the
internal degrees of freedom of the anti-$G_i$-particles associated
with $\{{\Bold{e}_{l_i}^{(\sBold{R}_i)}}\}$ since they are
characterized by opposite quantum numbers.

The analysis in this subsection has yielded two insights that may
be useful in model building. First, the Lie algebra possesses a
special basis that is particularly suited to model gauge bosons
and their physical attributes. Second, the matter field irreps for
the direct product group $G=\otimes G_n$ should include all
relevant combinations of irreps of the subgroups $G_i$. This leads to all combinations of internal quantum numbers.

\subsection{Dynamical quantum numbers}\label{sec. dynamics}

Since local internal symmetries can be used to model some
attributes of elementary particles, it is natural to include them
in dynamical models. The method has been known for a long time:
define a covariant derivative that facilitates building an
invariant Lagrangian density. The covariant derivative encodes the
interactions and, hence, the dynamics associated with the internal
symmetries.

Consider a principal fiber bundle with structure group $G_i$ and
Minkowski space-time base space. Let
$\mathcal{A}_i(x):=\Bold{A}^{a_i}(x)\otimes\bs{g}_{a_i}$ be the
local coordinate expression on the base space of the gauge
potential (the pull-back under a local trivialization of the
connection defined on the principal bundle). $\Bold{A}^{a_i}(x)$
is a complex one-form on the base space whose hermitian
components $A_{\mu}^{a_i}(x)$ represent gauge fields. The gauge
field self-interactions are encoded in the covariant derivative of
the gauge potential
\begin{equation}\label{field strength}
  \mathcal{F}_i(x):=D\mathcal{A}_i(x)=d\mathcal{A}_i(x)
  +\tfrac{1}{2}[\mathcal{A}_i(x),\mathcal{A}_i(x)]
  =:\Bold{F}^{a_i}(x)\bs{g}_{a_i}
\end{equation}
where $\Bold{F}^{a_i}$ is a two-form on the base space. In the
special basis determined by the decomposition of the previous
section, the commutator term describes interactions between gauge
fields characterized by the kinematical quantum numbers
$\Bold{q}_i$ by virtue of (\ref{eigenvalues}).

Matter fields will be sections of a tensor product bundle
$S\otimes V$. Here $S$ is a spinor bundle over space-time with
typical fiber $\C^4$, and $V$ is a vector bundle associated to the
gauge principal bundle with typical fiber
$V_{\times\sBold{R}_n}:=\otimes V_{\sBold{R}_i}$.

A basis element in $\C^4\otimes V_{\times\sBold{R}_n}$ will be
denoted $\Bold{e}_{\times l_n}^{(\times\sBold{R}_n)}
:=\otimes\Bold{e}_{l_i}^{(\sBold{R}_i)}$. (For clarity, we will
not make the spinor index explicit.) Vector space
$V_{\times\sBold{R}_n}$ furnishes the representation
$\rho{{^{(\times \sBold{R}_n)}}}(\otimes G_n)$. It is this
representation that determines the gauge/matter field interactions
via the covariant derivative $\D$;
\begin{equation}\label{matter field covariant derivative}
  \D\Bold{\Psi}^{(\times\sBold{R}_n)}(x)=\left[\not\!{\partial}
  +\rho_e{'{^{(\times \sBold{R}_n)}}}(\not\!\!\mathcal{A})
  \right]\Bold{\Psi}^{(\times\sBold{R}_n)}(x)
\end{equation}
where $\not\!\!\mathcal{A}:=
\mathrm{i}_{\gamma}\mathcal{A}\in\oplus\mathcal{G}_n$ and
$\Bold{\Psi}^{(\times\sBold{R}_n)}(x):=\mathit{\Psi}^{\times
l_n}(x)\Bold{e}_{\times l_n}^{(\times\sBold{R}_n)}$.

There is a scale ambiguity that resides in the matter field
covariant derivative. The inner product on
$\rho_e'(\mathcal{G}_i)$ for any \emph{faithful} representation of
a simple or abelian $G_i$ is proportional to the inner
product on $\mathcal{G}_i$. This implies the matrices in the
covariant derivative (\ref{matter field covariant derivative}) are
determined only up to overall constants $\kappa_{\mathcal{G}_i}$
---relative to the scale of the gauge fields. These constants are
conventionally interpreted as coupling constants characterizing
the matter field/gauge boson interaction. We choose the coupling
constants so that, given gauge and matter field normalizations,
the parameters in the matter field covariant derivative that
characterize neutral gauge/matter field interactions coincide with
the kinematical quantum numbers $\Bold{q}_{s_i}^{(m)}$.

With this choice, \emph{the parameters characterizing couplings in
both the gauge and matter field covariant derivatives are the
kinematical quantum numbers}.

The (bare) Lagrangian density is comprised of the usual Yang-Mills
terms, spinor matter field terms, ghost terms, and gauge fixing
terms. The Yang-Mills terms are
\begin{equation}\label{Yang-Mills}
  -\frac{1}{4}\sum_i\kappa_i\mathcal{F}_i\cdot\mathcal{F}_i
\end{equation}
where $\kappa_i$ are positive real constants. The dot product
represents the Minkowski metric and an $Ad(g_i)$ invariant inner
product on each $\mathcal{G}_i$. Normalization of the Lie algebra
basis elements is determined by a choice of Lie algebra inner
product. Since the $\mathcal{G}_i$ are compact simple and/or
$u(1)$ subalgebras, the inner product on each
subspace\footnote{Since the Lie algebra is a direct sum, the inner
product between subspaces vanishes.} $\{\bs{g}_{a_i}\}$ is
classified by a single constant; and, hence, the normalization can
be conveniently fixed by taking
$\bs{g}_{a_i}\cdot\bs{g}_{b_i}=\kappa_i^{-1}\delta_{a_i,b_i}$.
This normalization effectively fixes the scale of
$\Bold{A}^{a_i}(x)$ and hence also the gauge fields
$A_{\mu}^{a_i}(x)$ given the standard Minkowski inner product.

The most general spinor matter field Lagrangian density consistent
with the requisite symmetries is, according to the suggestion from
the previous section, comprised of a sum over all the inequivalent
faithful irreps of the elementary matter fields:
\begin{equation}\label{matter lagrangian}
  \mathcal{L}_m
  =i\sum_{\times\sBold{R}_n}\kappa_{\times\sBold{R}_n}
  \overline{\Bold{\Psi}}^{(\times\sBold{R}_n)}\cdot
  \D\Bold{\Psi}^{(\times\sBold{R}_n)}+\mbox{mass terms}
\end{equation}
where $\overline{\Bold{\Psi}}^{(\times\sBold{R}_n)}$ is a section
of the conjugate bundle $\overline{S\otimes V}=\overline{S}\otimes
\overline{V}$ and $\kappa_{\times\sBold{R}_n}$ are positive real
constants that are constrained by various consistency conditions;
for example, anomaly considerations and CPT symmetry. It is clear
that $\delta\mathcal{L}_m=0$ for $\Bold{\Psi}(x)\rightarrow
\exp\{\theta(x)^{a_i}\rho_e{'}(\bs{g}_{a_i})\}\Bold{\Psi}(x)$
despite the presence of $\kappa_{\times\sBold{R}_n}$ (assuming
appropriate mass terms).

The dot product here represents a Lorentz and $\rho(g)$ invariant
hermitian matter field inner product. It is not true in general
that $\kappa_{\times\sBold{R}_n}$ can be absorbed by a choice of
matter field inner product: If the matter fields are functionally
related, the associated factors cannot be absorbed by a choice of
inner product (equivalently by a field redefinition). For example,
suppose some components of two matter fields, say $\Bold{\Psi}^{(\times\sBold{R}_n)}$
and $\Bold{\Psi}^{(\times\widetilde{\sBold{R}}_n)}$, are related
by $(x,\Bold{\Psi}_x^{(\times\sBold{R}_i)})\in S\otimes V_{\sBold{R}_i}$ and
$(x,\Bold{\Psi}_x^{(\times\widetilde{\sBold{R}}_i)})\in S\otimes
\overline{V_{\sBold{R}_i}}$ in a given chart and trivialization. In other words, the $\widetilde{\Bold{R}}_i$ representation is the conjugate representation of $\Bold{R}_i$. If the representation is not real, the ratio
$\kappa_{\times\sBold{R}_n}/\kappa_{\times\widetilde{\sBold{R}}_n}$
may be non-trivial because the inner products can't be adjusted separately. This
persists even after renormalization. Note that for $n=1$ or if the
fields are not functionally related, the
$\kappa_{\times\sBold{R}_n}$ can always be absorbed into the scale
of $\Bold{e}_{\times l_n}^{(\times\sBold{R}_n)}$, and so \emph{the
$\kappa_{\times\sBold{R}_n}$ can be non-trivial only for direct
product groups with functionally related matter fields}. The
possibility of non-trivial factors $\kappa_{\times\sBold{R}_n}$ in
the matter field Lagrangian density is a key element in our
analysis.

For each individual subgroup $G_i$, the gauge and matter field
terms in the Lagrangian density give rise to the conserved
currents
\begin{equation}\label{conserved current 1}
J_{(a_i)}^{\mu}=
-\mathcal{F}_i^{\mu\nu}\cdot\left[\bs{g}_{a_i},{\mathcal{A}_i}_{\nu}\right]
  +j_{(a_i)}^{\mu}
\end{equation}
where
\begin{equation}\label{conserved current 2}
  j_{(a_i)}^{\mu}=\sum_{\times\sBold{R}_n}
  \kappa_{\times\sBold{R}_n}
  \overline{\Bold{\Psi}}^{(\times\sBold{R}_n)}
  \cdot\gamma^{\mu}\rho_e{'{^{(\times\sBold{R}_n)}}}
  (\bs{g}_{a_{i}})\Bold{\Psi}^{(\times\sBold{R}_n)}
\end{equation}
are the covariantly conserved matter field currents. In
particular, the neutral conserved currents  associated with $G_i$
are
\begin{eqnarray}\label{conserved current 3}
  J_{(s_i)}^{\mu}&=&
-\mathcal{F}_i^{\mu\nu}\cdot\left[\bs{h}_{s_i},
{\mathcal{A}_i}_{\nu}\right]
  +j_{(s_i)}^{\mu}\notag\\
  &=&-q_{s_i}F_{-\alpha_{k_i}}^{\mu\nu}A^{\alpha_{k_i}}_{\nu}
  +\sum_{\times\sBold{R}_n}\kappa_{\times\sBold{R}_n}
  (\Bold{q}^{(m)}_{s_i})
  \overline{\Bold{\Psi}}^{(\times\sBold{R}_n)}
  \cdot\gamma^{\mu}\Bold{\Psi}^{(\times\sBold{R}_n)}\,.
\end{eqnarray}
The constants $\kappa_{\times\sBold{R}_n}(\Bold{q}^{(m)}_{s_i})$
will be termed \emph{coupling strengths} since they represent the
scale of the gauge/matter field couplings given matter field
normalizations. Evidently, not all matter field currents
contribute to interactions on an equal basis. This is significant
because particles characterized by a set of internal quantum
numbers will appear to have scaled  internal quantum numbers  when
interacting with gauge bosons.

However, in order to conclude this, we must first confirm that the
normalization freedom in the Lagrangian density allows us to
maintain equality between the \emph{renormalized} parameters
$\Bold{q}_{i}$ and $\Bold{q}^{(m)}_{s_i}$ appearing in equations
(\ref{charge operators}) and the kinematical quantum numbers.
Moreover, we must verify that the $\kappa_{\times\sBold{R}_n}$ do
not destroy the assumed local symmetries.

The associated neutral quantum charge operators $Q_{(s_i)}:=-i\int
J^0_{(s_i)}dV$ of currents (\ref{conserved current 3}) encode the
\emph{dynamical (internal) quantum numbers} in the sense that
\begin{eqnarray}\label{charge operators}
  &&[Q_{(s_i)},A^{\alpha_{k_j}}_{\perp}]
  =q_{s_i}A^{\alpha_{k_j}}_{\perp}\delta_{ij}\notag\\
   &&[Q_{(s_i)},\Bold{\Psi}^{(\times\sBold{R}_n)}]
   =\Bold{q}^{(m)}_{s_i}\Bold{\Psi}^{(\times\sBold{R}_n)}
\end{eqnarray}
where the gauge and matter fields have been promoted to quantum
operators and $A^{\alpha_{k_i}}_{\perp}$ are the transverse gauge
fields. The second relation follows because the conjugate momentum
of $\Bold{\Psi}^{(\times\sBold{R}_n)}$ is
$\kappa_{\times\sBold{R}_n}\overline{\Bold{\Psi}}^{(\times\sBold{R}_n)}$
as determined from (\ref{matter lagrangian}).

Equations (\ref{charge operators}) are in terms of bare
quantities, but they are required to be valid for renormalized quantities
as well. In particular, \emph{the renormalized dynamical quantum
numbers are required to coincide with the kinematical quantum
numbers}. Under the renormalizations
\begin{equation}\label{renormalization 1}
\mathcal{A}_{i}^{\mathrm{B}}\rightarrow
Z_{\mathcal{A}_{i}}^{1/2}\mathcal{A}_{i}^{\mathrm{R}}
\end{equation}
and
\begin{equation}\label{renormalization 2}
{\Bold{\Psi}^{(\times\sBold{R}_n)}}^{\mathrm{B}}\rightarrow
Z_{\sBold{\Psi}^{(\times\ssBold{R}_n)}}^{1/2}{\Bold{\Psi}^{(\times\sBold{R}_n)}}^{\mathrm{R}}\;,
\end{equation}
the basis elements $\bs{g}_{a_i}$ (equivalently the $\kappa_i$)
can be re-scaled so that
$\Bold{q}_i^{\mathrm{B}}=Z_{\mathcal{A}_i}^{-1/2}\Bold{q}_i^{\mathrm{R}}$.
Likewise, the coupling constants $\kappa_{\mathcal{G}_i}$ can be
chosen so that ${\Bold{q}_{s_i}^{(m)}}^{\mathrm{B}}
=Z_{\mathcal{A}_i}^{-1/2}{\Bold{q}_{s_i}^{(m)}}^{\mathrm{R}}$.
Consequently the relations (\ref{charge operators}) will be
maintained under renormalization. The renormalized form of
equations (\ref{charge operators}) are to be compared to
(\ref{eigenvalue}) and (\ref{full representation}). That they are
consistent is a consequence of: (i) the covariant derivatives
(\ref{field strength}) and (\ref{matter field covariant
derivative}), (ii) our choice of Lie algebra inner product, and
(iii) our choice of coupling constants. This consistency ensures
the renormalized gauge and matter fields appearing in the
Lagrangian density are the elementary fields associated with the
quantum numbers $\Bold{q}_{i}$ and $\Bold{q}_{s_i}^{(m)}$. It
should be emphasized that the coupling constants are implicit in
$\Bold{q}_{i}$ and $\Bold{q}_{s_i}^{(m)}$, and the
$\kappa_{\times\sBold{R}_n}$ do not get renormalized; or, rather,
non-trivial $\kappa_{\times\sBold{R}_n}$ persist after
renormalization of $\Bold{\Psi}^{(\times\sBold{R}_n)}$.

We will refer to the two equivalent types of quantum
numbers --- renormalized dynamical quantum numbers and kinematical
quantum numbers --- by the common term \emph{intrinsic charges}. The
renormalized coupling strengths
$\kappa_{\times\sBold{R}_n}({\Bold{q}_{s_i}^{(m)}}^\mathrm{R})$ in
the renormalized currents (\ref{conserved current 3}) will be
called \emph{extrinsic charges}.

To maintain the assumed local symmetries of the Lagrangian
density, $Q_{(a_i)}$ and $\{\bs{g}_{a_i}\}$, along with their
associated commutation relations, must determine isometric
algebras. We readily find
\begin{eqnarray}\label{current algebra}
  [J^0_{(a_i)},J^0_{(b_j)}]&=&\delta_{ij}C^{c_j}_{a_ib_j}
\left\{-\mathcal{F}_i^{\mu\nu}\cdot\left[\bs{g}_{c_j},
{\mathcal{A}_i}_{\nu}\right]
 +\sum_{\times\sBold{R}_n}
  \kappa_{\times\sBold{R}_n}
  \overline{\Bold{\Psi}}^{(\times\sBold{R}_n)}
  \rho_e{'{^{(\times\sBold{R}_n)}}}
  (\bs{g}_{c_j})\Bold{\Psi}^{(\times\sBold{R}_n)}\right\}\notag\\\notag\\
  &= & \delta_{ij}C^{c_j}_{a_ib_j}J^0_{(c_j)}
\end{eqnarray}
where $C_{a_ib_i}^{c_i}$ are the structure constants of
$\mathcal{G}_i$.

\emph{It is crucial that the $\kappa_{\times\sBold{R}_n}$ factors
do not spoil the equality between the kinematical and dynamical
internal quantum numbers or the local symmetries}.

The analysis in this subsection leads to the conclusion that,
\emph{in some cases at least, the intrinsic charges of matter
fields do not fully determine their coupling strengths to gauge
bosons}. Stated otherwise, the intrinsic and extrinsic charges of matter fields are not necessariy equivalent.

\section{An Example} \label{sec. example}
It is useful to illustrate the generalities of the previous
section with a concrete example. We choose the product group
$SU(3)\otimes U(2)$ for obvious reasons. The associated Lie
algebra is $su(3)\oplus u(2)\cong su(3)\oplus su(2)\oplus u(1)$.

The decomposition of $su(3)$ with respect to its Cartan subalgebra
is well-known and need not be reproduced here. Suffice it to say
that it contains two neutral bosons, two bosons characterized by a
single non-zero quantum number, and the remaining four bosons
characterized by two quantum numbers. They come in pairs with
opposite charges.
\vspace{.25in}

\noindent\textbf{Remark}: This decomposition has interesting
implications for QCD. The conventional view is that all eight
gluons carry color charges. Our point-of-view differs
substantially; indeed, from our standpoint there are two neutral
gluons (with concomitant neutral QCD currents). In particular,
this suggests that the salient feature of color confinement is
related to the \emph{representation} and not the \emph{color
charge}. That is, presumably, QCD gauge bosons are confined not
because they carry color charge but because they furnish the
adjoint representation
of $SU_C(3)$.\footnote{From this perspective, $U(3)$ would not
imply a new long-range massless gauge boson, and we postulate that
$U(3)$ is the symmetry group for the strong interaction. There are
good reasons which will be presented elsewhere to advocate
$U(3)$ as the symmetry group of strong interactions.}

\vspace{.25in}

The decomposition of $u(2)$ is
\begin{equation}\label{u2 decomposition}
[\bs{h}_{s},\bs{h}_{r}]=0
\end{equation}
and
\begin{equation}
  [\bs{h}_{s},\bs{e}_{\pm}]=\pm q_{s}\bs{e}_{\pm}\;.
\end{equation}
where $r,s\in\{1,2\}$. It follows that there are two neutral
bosons and two oppositely charged bosons characterized by two
quantum numbers.

We will consider only Dirac matter fields in the fundamental
representation of $SU(3)$ and $U(2)$. Consequently, the matter
fields are sections of an associated fiber bundle with typical
fiber $\C^4\otimes\C^3\otimes\C^2$. Since $su(3)$ and $u(2)$ are
both rank two algebras, these matter fields can be labelled by
four internal quantum numbers; two associated with $SU(3)$ and two
with $U(2)$. According to the remark at the end of section
\ref{sec. kinematics}, inequivalent irreps of the direct product
group are postulated to include the $(\bs{3},\bs{2})$ and
$(\bs{3},\overline{\bs{2}})$ and their anti-fields
$(\overline{\bs{3}},\overline{\bs{2}})$ and
$(\overline{\bs{3}},\bs{2})$.

The field furnishing the $(\bs{3},\bs{2})$ is a section of the
bundle $S\otimes V_{SU(3)}\otimes V_{U(2)}$. Let $\{\Bold{e}_A\}$
span $V_{\bs{3}}\cong\mathbb{C}^3$ and $\{\Bold{e}_a\}$ span
$V_{\bs{2}}\cong\mathbb{C}^2$. Internal degrees of freedom of
elementary matter fields are associated with the basis
$\{\Bold{e}_{Aa}\}:=\{\Bold{e}_A\otimes\Bold{e}_a\}$ that spans
$\C^3\otimes\C^2$. Explicitly, given a trivialization and
coordinate chart, the matter field is
$\Bold{\Psi}=\mathit{\Psi}^{Aa}\Bold{e}_{Aa}$ with the spinor
index implicit. The most general two-dimensional representation
of $u(2)$ furnished by $\Bold{\Psi}$ consistent with decomposition
(\ref{u2 decomposition}) is given by
\begin{equation} \rho_e{'}(\bs{h}_1)=i\left(\begin{array}{cc}
  R & 0 \\
  0 & S
\end{array}\right),\;
\rho_e{'}(\bs{e}_+)=i\left(\begin{array}{cc}
  0 & T \\
  0 & 0
\end{array}\right),\;
\rho_e{'}(\bs{e}_-)=i\left(\begin{array}{cc}
  0 & 0 \\
  T & 0
\end{array}\right),\;
\rho_e{'}(\bs{h}_2)=i\left(\begin{array}{cc}
  U & 0 \\
  0 & V
\end{array}\right),\;
\end{equation}
where $R,S,T,U,V$ are real constants. Evidently, the elementary
matter fields $\mathit{\Psi}_{A1}$ and $\mathit{\Psi}_{A2}$ have
$U(2)$ kinematical quantum numbers $(R,U)$ and $(S,V)$
respectively. The choice of Lie algebra inner product and the
orthogonality of the basis elements determine to some extent the
real constants. Similarly, the three-dimensional representation
can be derived based on the analogous decomposition of $SU(3)$.

The $(\bs{3},\overline{\bs{2}})$ field $\widetilde{\Bold{\Psi}}$
is a section of $S\otimes V_{SU(3)}\otimes
V_{\overline{U(2)}}=S\otimes V_{SU(3)}\otimes
\overline{V_{U(2)}}$. Bundle $S\otimes V_{SU(3)}\otimes
\overline{V_{U(2)}}$ is the image under the bundle morphism
\begin{eqnarray}
F:S\otimes V_{SU(3)}\otimes V_{U(2)}&\longrightarrow& S\otimes
V_{SU(3)}\otimes \overline{V_{U(2)}}\notag\\
\rule{0in}{.2in}
(x,\mathit{\Psi}^{Aa}(x)\Bold{e}_{Aa})&\longmapsto&
(x,[i\tau_2]^{\bar{a}}_a\mathit{\Psi}^{Aa}(x)(\Bold{e}_A\otimes
\Bold{e}_a^*))
=:(x,\mathit{\Psi}^{A\bar{a}}(x)\Bold{e}_{A\bar{a}})
\end{eqnarray}
in a given chart and trivialization. The corresponding conjugate
$u(2)$ representation is
$\overline{\rho}_e{'}:=(i\tau_2)^{\dag}(\rho_e{'})^{*}(i\tau_2)$.
Then, in particular,
\begin{equation}
\overline{\rho}_e{'}(\bs{h}_s)\widetilde{\Bold{\Psi}}
=(i\tau_2)\rho_e{'}(\bs{h}_s)^{*}\mathit{\Psi}^{Aa}(\Bold{e}_A\otimes
\Bold{e}_a^*)
=-q^{(m)}_{s,a}(i\tau_2)\mathit{\Psi}^{Aa}(\Bold{e}_A\otimes
\Bold{e}_a^*) =-\Bold{q}^{(m)}_{s}\widetilde{\Bold{\Psi}}\;.
\end{equation}
So $\widetilde{\Bold{\Psi}}$ indeed transforms by the conjugate
representation of $U(2)$. (We emphasize that there is no
conjugation associated with the $SU(3)$ or Dirac index here.)

The covariant derivative acting on the matter fields in the
$(\bs{3},\bs{2})$ and $(\bs{3},\overline{\bs{2}})$ representation
is
\begin{equation}\label{covariant derivative 1}
  (\D\Bold{\Psi})=\left[\not\!{\partial}[\bs{1}]^{Aa}_{Bb}
  +\not\!G^{\alpha}\,[\bs{\Lambda}_{\alpha}]^A_B\otimes [\bs{1}]^{a}_{b}
  +[\bs{1}]^{A}_{B}\;\otimes\not \!g^{\sigma}\, [\Bold{\lambda}_{\sigma}]^a_b\;\right]
  \mathit{\Psi}^{Bb}\Bold{e}_{Aa}
\end{equation}
and
\begin{equation}\label{covariant derivative 2}
  (\widetilde{\D}\widetilde{\Bold{\Psi}})=\left[\not\!{\partial}[\bs{1}]^{A\bar{a}}_{B\bar{b}}
  +\not\!G^{\alpha}\,[\bs{\Lambda}_{\alpha}]^A_B\otimes [\bs{1}]^{\bar{a}}_{\bar{b}}
  +[\bs{1}]^{A}_{B}\;\otimes\not {\!g^{\sigma}}^*\, [\overline{\Bold{\lambda}}_{\sigma}]^{\bar{a}}_{\bar{b}}
  \;\right]
  \mathit{\Psi}^{B\bar{b}}\Bold{e}_{A\bar{a}}
\end{equation}
respectively. These yield the matter field Lagrangian density;
\begin{eqnarray}\label{matter field Lagrangian}
  \mathcal{L}_m&=&i\kappa\overline{\mathit{\Psi}^{A'a'}}\left[\not\!{\partial}[\bs{1}]^{Aa}_{Bb}
  +\not\!G^{\alpha}\,[\bs{\Lambda}_{\alpha}]^A_B\otimes [\bs{1}]^{a}_{b}
  +\not \!g^{\sigma}\,[\bs{1}]^{A}_{B}\otimes [\Bold{\lambda}_{\sigma}]^a_b\;\right]
  \mathit{\Psi}^{Bb}\delta_{A'A}\delta_{a'a}\notag\\
  &&+i\widetilde{\kappa}\overline{\mathit{\Psi}^{A'\bar{a}'}}\left[\not\!{\partial}[\bs{1}]^{A\bar{a}}_{B\bar{b}}
  +\not\!G^{\alpha}\,[\bs{\Lambda}_{\alpha}]^A_B\otimes [\bs{1}]^{\bar{a}}_{\bar{b}}
  +\not {\!g^{\sigma}}^*\,[\bs{1}]^{A}_{B}\otimes [\overline{\Bold{\lambda}}_{\sigma}]^{\bar{a}}_{\bar{b}}
  \;\right]
  \mathit{\Psi}^{B\bar{b}}\delta_{A'A}\delta_{\bar{a}'\bar{a}}\notag\\
  &&+\mbox{mass terms}\;.
\end{eqnarray}
Note that it is not possible to absorb both $\kappa$ and
$\widetilde{\kappa}$ by a field redefinition because
$\mathit{\Psi}^{A\bar{a}}=[i\tau_2]_a^{\bar{a}}\mathit{\Psi}^{Aa}$
and $\Bold{e}_a\cdot\Bold{e}_a=\Bold{e}^*_a\cdot\Bold{e}^*_a$;
implying that the ratio $\widetilde{\kappa}/\kappa$ is non-trivial
in this example.

The corresponding $U(2)$ and $SU(3)$ currents are
\begin{equation}\label{currents 1}
  j^{\mu}_{(\sigma)}=\kappa\overline{\mathit{\Psi}_a^{A}}\gamma^{\mu}
  [\Bold{\lambda}_{\sigma}]^a_b
  \mathit{\Psi}_A^{b}
  +\widetilde{\kappa}\overline{\mathit{\Psi}^A_{\bar{a}}}
  \gamma^{\mu}[\overline{\Bold{\lambda}}_{\sigma}]^{\bar{a}}_{\bar{b}}
  \mathit{\Psi}_A^{\bar{b}}
\end{equation}
and
\begin{equation}\label{currents 2}
  j^{\mu}_{(\alpha)}=\kappa\overline{\mathit{\Psi}_A^{a}}\gamma^{\mu}
  [\bs{\Lambda}_{\alpha}]^A_B
  \mathit{\Psi}_a^{B}
  +\widetilde{\kappa}\overline{\mathit{\Psi}_A^{\bar{a}}}
  \gamma^{\mu}[\bs{\Lambda}_{\alpha}]^{A}_{B}\mathit{\Psi}^B_{\bar{a}}
  =(\kappa+\widetilde{\kappa})\overline{\mathit{\Psi}_A^{a}}
  \gamma^{\mu}[\bs{\Lambda}_{\alpha}]^{A}_{B}\mathit{\Psi}_a^{B}
\end{equation}
respectively. Evidently, if $(\kappa+\widetilde{\kappa})=1$ the
original $SU(3)$ coupling strength is preserved, i.e., the $SU(3)$
intrinsic and extrinsic charges are equivalent. However, in this
case, the $U(2)$ external charges are fractional relative to the
internal charges since $\kappa,\widetilde{\kappa}\neq 0$ by
assumption. In a realistic model, the ratio is ultimately fixed
by anomaly considerations.

\section{Conclusions}
We have studied the relation between quantum numbers and coupling
strengths for internal symmetry groups that are direct product
groups. It was argued that the Lagrangian density can have
non-trivial factors multiplying matter field terms that do not
spoil the local invariance or the equality between kinematical and
dynamical quantum numbers. However, these non-trivial factors do
affect the coupling strengths of gauge/matter field interactions.
This relationship is summarized by the statement that intrinsic
and extrinsic charges are not necessarily equivalent.

In our specific example, we found that non-trivial factors in the
Lagrangian density can preserve the $SU(3)$ coupling strength if
the factors sum to unity. On the other hand, the $U(2)$ coupling
strengths are modified. The example shows that elementary
particles can have intrinsic $U(2)$ charges that, nevertheless,
appear to be fractional extrinsic charges due to dynamics. This
has obvious implications for the Standard Model and will be
explored further in a separate paper.

% ------------------------------------------------------------------------

\end{document}